# Spin Seebeck Power Conversion

Adam B. Cahaya[1], Oleg A. Tretiakov[1], and G. E. W. Bauer[1,2,3]

[1]Institute for Materials Research, Tohoku University, Aoba-ku, Katahira 2-1-1, Sendai 980-8577, Japan,
[2]WPI Advanced Institute for Materials Research, Tohoku University, Aoba-ku, Katahira 2-1-1, Sendai 980-8577, Japan,
[3]Kavli Institute of NanoScience, TU Delft Lorentzweg 1, 2628 CJ Delft, The Netherlands

Spin caloritronics is the science and technology to control spin, charge, and heat currents in magnetic nanostructures. The spin degree of freedom provides new strategies for thermolelectric power generation that have not yet been fully explored. After an elementary introduction into conventional thermoelectrics and spintronics, we give a brief review of the physics of spin caloritronics. We discuss spin-dependent thermoelectrics based on the the two-current model in metallic magnets as well as the spin Seebeck and Peltier effects that are based on spin wave excitations in ferromagnets. We derive expressions for the efficiency and figure of merit $ZT$ of several spin caloritronic devices.

*Index Terms*—spin Seebeck, spin caloritronics, thermoelectricity

## I. Introduction

Thermoelectric phenomena can transform heat currents into electric power and vice versa. The Seebeck effect refers to the generation of an electromotive force by a temperature gradient [46], which can be employed for thermoelectric power generators that convert waste heat into useful electric energy. The efficiency of these generators can be parameterized by the dimensionless figure of merit $ZT$ [46]. The increased Ohmic losses and Joule heating that are associated with decreasing feature size of the elements in integrated circuits will lead to a breakdown of Moore's law in the near future. Strategies to delay this threat include thermoelectrics that while not reducing at least help to manage excess heat. Enhanced transport properties of nanostructures promise reprieve [14], [18]. However, in spite of some progress in search for new materials the efficiency of thermoelectric power generators appears to have been stalled [64] (although a more optimistic view has been expressed as well [24]). Spin caloritronics, in which the spin degree of freedom is employed, is one direction to look for new strategies [4]. Huge Peltier effects in constantan nanopillars [53] and Seebeck coefficients in tunneling junctions [37] have been reported.

Even more promising is the recently discovered so-called spin Seebeck effect [61], especially in the "longitudinal" configuration based on the magnetic insulator yttrium iron garnet (YIG) [59]. The spin Seebeck effect converts a temperature difference between a ferromagnet and a normal metal contact into electric energy by pumping a spin current into the normal metal (such as Pt) that by the inverse spin Hall effect is converted into a charge current [68]. Large-area applications are being considered by the Japanese electronics maker NEC since thermoelectric power can be generated by simply coating in principle arbitrary materials [35].

Here we review the physics of the spin-dependent and spin Seebeck effects with an emphasis on estimating the power efficiency. We also propose a new configuration to harvest a thermally generated spin current by a ferromagnetic metal instead of the spin Hall effect. The latter turns out to be advantageous for thermoelectric power generation with a large figure of merit and higher output voltages for small structures [8]. We do not discuss here the thermoelectric properties of magnetic tunnel junctions, such as the tunneling magneto Seebeck effect and related properties [65].

This article is organized as follows. In Section II, we briefly recapitulate the principles of conventional thermoelectrics including the efficiency and figure of merit $ZT$ of thermocouples. In Section III, we review (and emphasize the difference between) the spin-dependent Seebeck effect and the spin Seebeck effect. We analyze the spin-heat transport by thermal spin pumping, i.e. the physical origin of the spin Seebeck effect. In Section III-C, we describe spin-heat-charge conversion utilizing the spin Hall effect and discuss thermoelectric coatings [35]. In Section III-D we suggest an alternative spin Seebeck power generating device that is based on the spin-filtering effect of metallic ferromagnets. Since we cannot naively use the conventional definition of the figure of merit, we derive in Section IV new ones for generators based on the spin Seebeck effect for the two different spin to charge current conversion schemes. We find that the spin valve based on spin Seebeck effect has a much higher efficiency, figure of merit, and an output voltage that does not decrease with sample size. We finish with conclusions and an outlook.

## II. Thermoelectric Power Conversion

### A. Seebeck effect

Named after their discoverers, the *Seebeck effect* describes the generation of a thermovoltage gradient $\nabla V$ (the electron charge is $-e$ therefore $e > 0$) by a temperature gradient $\nabla T$, while the *Peltier effect* refers to the heat current associated with a electric current [46] These are the basic thermoelectric phenomena that govern the conversion of heat into electric energy and vice versa in electrical conductors. The Peltier effect expresses the heat current $\mathbf{J}_Q$ in the presence of an electric current $\mathbf{J}_c$, defined here as charge flow, i.e., opposite

to the particle current:

$$(\nabla V)^{\text{Seebeck}} = S\nabla T, \tag{1}$$

$$\mathbf{J}_Q^{\text{Peltier}} = \Pi\mathbf{J}_c, \tag{2}$$

where $S \equiv (\nabla V/\nabla T)_{J_c=0}$ and $\Pi \equiv (J_Q/J_c)_{\Delta V=0}$ are the Seebeck and Peltier coefficients, respectively. When heat and electric currents are uniform over a cross section $A$, the current vectors $\mathbf{J}_x = A\mathbf{j}_x$ are described in terms of the current densities $\mathbf{j}_x$. By combining the Seebeck effect with Ohm's law $\mathbf{J}_c = -\sigma A\nabla V$, and Peltier effect with Fourier's law $\mathbf{J}_Q = -\kappa A\nabla T$, where $\sigma$ and $\kappa$ are the electric (at $\nabla T = 0$) and thermal (at $\mathbf{J}_c = 0$) conductivities, the traditional thermoelectric relations read:

$$\mathbf{J}_c = -\sigma A\left(\nabla V + S\nabla T\right), \tag{3}$$

$$\mathbf{J}_Q = \Pi\mathbf{J}_c - \kappa A\nabla T, \tag{4}$$

or in matrix form:

$$\begin{pmatrix} -\nabla V \\ \mathbf{J}_Q \end{pmatrix} = \begin{pmatrix} 1/(\sigma A) & S \\ \Pi & -\kappa A \end{pmatrix} \begin{pmatrix} \mathbf{J}_c \\ \nabla T \end{pmatrix}. \tag{5}$$

The currents $\mathbf{J}_c$ and $\mathbf{J}_Q$ arise due to external $\nabla V$, $\nabla T$ which are referred to as "forces". The conjugate pairs of currents and forces can be balanced with respect to each other by demanding that the sum of their products equals the entropy production per unit volume $\dot{\mathcal{S}}$ [13], [41]:

$$T\dot{\mathcal{S}} = -\mathbf{J}_Q \cdot \frac{\nabla T}{T} - \mathbf{J}_c \cdot \nabla V. \tag{6}$$

When the electric current $\mathbf{J}_c$ is driven by $-\nabla V$ and the heat current $\mathbf{J}_Q$ by $-\nabla T/T$ we obtain the transformed equation

$$\begin{pmatrix} \mathbf{J}_c \\ \mathbf{J}_Q \end{pmatrix} = \sigma A \begin{pmatrix} 1 & ST \\ \Pi & (\kappa/\sigma + \Pi S)T \end{pmatrix} \begin{pmatrix} -\nabla V \\ -\nabla T/T \end{pmatrix}. \tag{7}$$

Onsager's reciprocity theorem [13], [41] now dictates that the response matrix must be symmetric, leading to the Onsager-Kelvin relation between Peltier and Seebeck coefficients $\Pi = ST$. Terms to second order in $S$ affect the heat conductance in the presence of charge currents. They are included since they significantly affect the thermoelectric efficiency, as show below.

In linear response, the Seebeck coefficient depends on the energy-dependent electric conductivity $\sigma(\varepsilon)$ as

$$S = -\frac{k_B}{e} \frac{\int d\varepsilon \frac{\partial f(\varepsilon)}{\partial \varepsilon} \sigma(\varepsilon) \frac{\varepsilon - \varepsilon_F}{k_B T}}{\int d\varepsilon \frac{\partial f(\varepsilon)}{\partial \varepsilon} \sigma(\varepsilon)}, \tag{8}$$

where $k_B$ is the Boltzmann constant, $f(\varepsilon)$ is the Fermi-Dirac distribution function, and $\varepsilon_F$ is the Fermi energy [10]. The leading order in the Sommerfeld expansion of $f(\varepsilon)$ near $\varepsilon_F$ gives Mott's law

$$S = -eL_0 T \frac{\partial_\varepsilon \sigma(\varepsilon)|_{\varepsilon_F}}{\sigma}, \tag{9}$$

where $L_0 = (\pi^2/3)(k_B/e)^2$ is the Lorenz constant and $\sigma = \sigma(\varepsilon_F)$. This approximation is valid when the conductivity varies slowly on the scale of the thermal window $k_B T$, which is often not the case at room temperature. The Seebeck coefficient of the free electron gas is negative since the density of states and conductivity increase with energy. The opposite holds when holes are the conducting carriers such as in p-doped semiconductors [71]. The Sommerfeld expansion also leads to the Wiedemann-Franz relation between the electronic part of the heat conductivity $\kappa_e$ and the electric conductivity:

$$\kappa_e = L_0 T\sigma. \tag{10}$$

Deviations from this relation are sometimes absorbed into a temperature dependent effective Lorenz constant.

### B. Thermocouple and thermopile

Consider a conductor as sketched in Fig. 1. At $z = 0$, the temperature is $T_H$, whereas at $z = l$ the temperature is $T_L$ and $T_H > T_L$. According to (4) the heat current $J_Q$ depends on the charge current $J_c$ and temperature gradient $dT/dz$. By current conservation and in the steady state $dJ_c/dz = 0$. The local heat balance therefore reads

$$\frac{dJ_Q}{dz} = -J_c \frac{dV}{dz} = \frac{J_c^2}{\sigma A} + SJ_c \frac{dT}{dz}, \tag{11}$$

where the last equality follows from (3). By substituting (11) into (4), we find

$$\frac{d^2T}{dz^2} = -\frac{J_c^2}{\sigma\kappa A^2}. \tag{12}$$

By double integration with boundary conditions $T(0) = T_H$ and $T(l) = T_L$, we obtain the temperature profile

$$T = -\frac{J_c^2}{2\sigma\kappa A^2}z^2 - \left(\frac{T_H - T_L}{l} - \frac{J_c^2 l}{2\sigma\kappa A^2}\right)z + T_H. \tag{13}$$

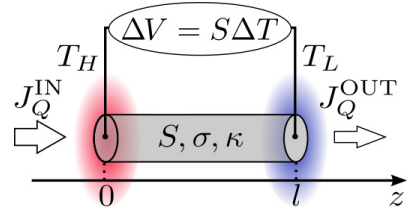

Fig. 1. Thermoelectric element consisting of a conducting wire whose ends are kept at different temperatures $T_H$ and $T_L$, generating a thermoelectric voltage $\Delta V = S\Delta T = S(T_H - T_L)$.

The heat currents at the terminals (see Fig. 1)

$$J_Q^{\text{IN}} = ST_H J_c + \kappa A \frac{T_H - T_L}{l} - \frac{J_c^2 l}{2\sigma A}, \tag{14}$$

$$J_Q^{\text{OUT}} = ST_L J_c + \kappa A \frac{T_H - T_L}{l} + \frac{J_c^2 l}{2\sigma A}, \tag{15}$$

contain a parabolic term in $J_c$, while the sum $J_Q^{\text{IN}} + J_Q^{\text{OUT}}$ depends linearly on $J_c$. The output voltage

$$\Delta V = V(l) - V(0) = S(T_H - T_L) - J_c l/\sigma A \tag{16}$$

$$= R_{\text{load}} J_c \tag{17}$$

when the current loop is closed by load resistance $R_{\text{load}}$ and

$$\Delta V = \frac{R_{\text{load}}}{R_{\text{load}} + \frac{l}{\sigma A}} S(T_H - T_L). \tag{18}$$

with $|\Delta V| < |S(T_H - T_L)|$. The thermoelectric voltage is suppressed by the Joule heating generated by the internal resistance $l/\sigma A$.

The differential thermoelectric relations (7) can now be rewritten in terms of voltage and temperature differences

$$\begin{pmatrix} J_c \\ \bar{J}_Q \end{pmatrix} = \frac{\sigma A}{l} \begin{pmatrix} 1 & S \\ ST & \kappa/\sigma + S^2 T \end{pmatrix} \begin{pmatrix} -\Delta V \\ T_H - T_L \end{pmatrix}. \quad (19)$$

Here we defined the average temperature $T = (T_H + T_L)/2$ and heat current $\bar{J}_Q = (J_Q^{\mathrm{IN}} + J_Q^{\mathrm{OUT}})/2$. The discrete representation is useful for nanostructures that limit the transport between large reservoirs or to model the discontinuities at heterointerfaces.

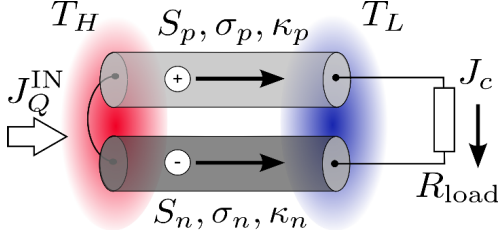

Fig. 2. A thermocouple consists of two conductors (for example $p$- and $n$-doped semiconductors) that are connected to a load resistance $R_{\mathrm{load}}$ under a temperature bias. The difference of Seebeck coefficients induces an isothermal voltage bias over the load resistance and drives a charge current through the circuit.

Next consider a device consisting of two conducting wires with Seebeck coefficients $S_p$ and $S_n$ connected on one side, i.e., a so-called thermocouple. The labels refer to semiconductors, where one can be p- and the other n-type doped such that $S_p > 0$ and $S_n < 0$. The circuit is closed by an external load resistor $R_{\mathrm{load}}$ (see Fig. 2). The two junctions are considered to have good thermal contact to heat reservoirs at temperatures $T_H$ and $T_L$. The difference in the Seebeck coefficients induces electromotive force $\varepsilon = (S_p - S_n)(T_H - T_L)$ and creates voltage $\Delta V$ over the load at constant temperature $T_L$. The induced current (see equivalent circuit in Fig. 3(a)) is $J_c = (S_p - S_n)(T_H - T_L)/(R_{\mathrm{load}} + R_p + R_n)$, where $R_{p(n)} = l_{p(n)}/(\sigma_{p(n)} A_{p(n)})$ is the internal resistance of the p(n)-type conductor. The work done by the system is the Ohmic dissipation

$$W = \left( \frac{(S_p - S_n)(T_H - T_L)}{R_{\mathrm{load}} + R_p + R_n} \right)^2 R_{\mathrm{load}} \quad (20)$$

in the load. This power depends on the signs of neither the temperature difference $\Delta T$ nor the difference in Seebeck coefficients. The total input heat current $J_Q^{\mathrm{IN}}$ at the contact to the hot bath is the sum of the heat flows into both conductors:

$$J_Q^{\mathrm{IN}} = (S_p - S_n) T_H J_c + (K_p + K_n)(T_H - T_L) - \frac{J_c^2}{2}(R_p + R_n),$$
$$(21)$$

where $K_{p(n)} = \kappa_{p(n)} A_{p(n)}/l_{p(n)}$ are the heat conductances.

The thermocouple is a heat engine as emphasized in Fig. 3 (b). The ratio of output power and input heat current is its efficiency $\eta = W/J_Q^{\mathrm{IN}}$. Defining $\Delta T = T_H - T_L (> 0)$, $S = S_p - S_n$, $K = K_p + K_n$, and $R_{\mathrm{int}} = R_p + R_n$, we find

$$\eta = \eta_C \left\{ 1 + \frac{R_{\mathrm{int}}}{R_{\mathrm{load}}} \frac{T}{T_H} \left[ 1 + \frac{R_{\mathrm{int}} K}{S^2 T} \left( 1 + \frac{R_{\mathrm{load}}}{R_{\mathrm{int}}} \right)^2 \right] \right\}^{-1}. \quad (22)$$

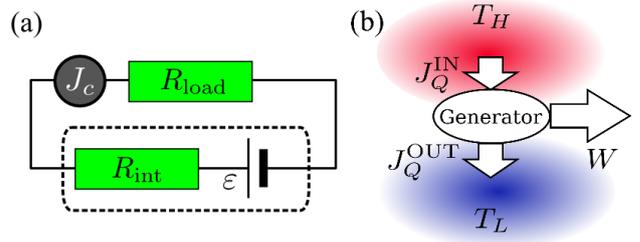

Fig. 3. (a) Equivalent circuit for a thermoelectric power generator, i.e., a battery that produces the electromotive force $\varepsilon = (S_p - S_n)(T_H - T_L)$ with internal resistance $R_{\mathrm{int}} = \rho_p l_p/A_p + \rho_n l_n/A_n$ that acts on an external load resistance $R_{\mathrm{load}}$. (b) A heat engine that generates power $W$ from an input heat current $J_Q^{\mathrm{IN}}$ has an efficiency $\eta = W/J_Q^{\mathrm{IN}} < 1$.

Here $\eta_C = \Delta T/T_H$ is the efficiency of a reversible Carnot cycle, which is the maximum thermodynamically allowed efficiency of any heat engine.

The output voltage can be increased by connecting thermocouples in series, as shown in Fig. 4, to form a "thermopile". The output voltage scales linear with the number of elements.

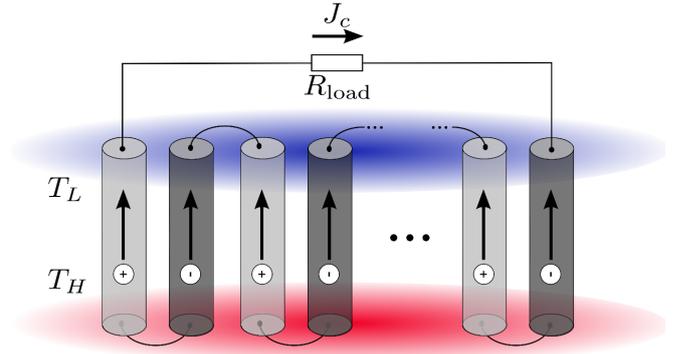

Fig. 4. A "thermopile" consisting of several thermocouples connected in series. The output voltage is proportional to the number of thermocouples.

### C. Figure of merit

The efficiency (22) of the thermocouple can be maximized by matching the load impedance to the internal one:

$$R_{\mathrm{load}}^{\mathrm{opt}} = R_{\mathrm{int}} \sqrt{1 + \frac{S^2 T}{R_{\mathrm{int}} K}}. \quad (23)$$

This leads to the optimized efficiency

$$\eta_{\max} = \frac{\Delta T}{T_H} \frac{\sqrt{1 + ZT} - 1}{\sqrt{1 + ZT} + 1 - \frac{\Delta T}{T_H}}, \quad (24)$$

where the dimensionless parameter

$$ZT = \frac{S^2 T}{R_{\mathrm{int}} K} = \frac{(S_p - S_n)^2 T}{\left( \kappa_p \frac{A_p}{l_p} + \kappa_n \frac{A_n}{l_n} \right) \left( \frac{l_p}{\sigma_p A_p} + \frac{l_n}{\sigma_n A_n} \right)} \quad (25)$$

is called the thermoelectric *figure of merit*. Note that $\eta_{\max}$ depends only on two parameters, $\Delta T/T_H$ and $ZT$. It is a monotonously increasing function of $ZT$ (Fig. 5) that parameterizes the dependence on the material parameters.

The output power of the optimized circuit in terms of $ZT$:

$$W_{\text{opt}} = \frac{\sqrt{1+ZT}}{\left(1+\sqrt{1+ZT}\right)^2} \frac{(S\Delta T)^2}{R_{\text{int}}}.$$ (26)

When the "power factor" $S^2/R_{\text{int}}$ is kept constant, $W_{\text{opt}}$ monotonically decreases with $ZT$. Thus, a high efficiency of thermoelectric conversion does not imply maximum output power. On the other hand, by rewriting

$$W_{\text{opt}} = \sqrt{1+ZT} \frac{\sqrt{1+ZT}-1}{\sqrt{1+ZT}+1} \frac{K(\Delta T)^2}{T},$$ (27)

it is seen to monotonically increase with $ZT$ when $K$ is kept constant. The largest possible "Carnot" efficiency is reached

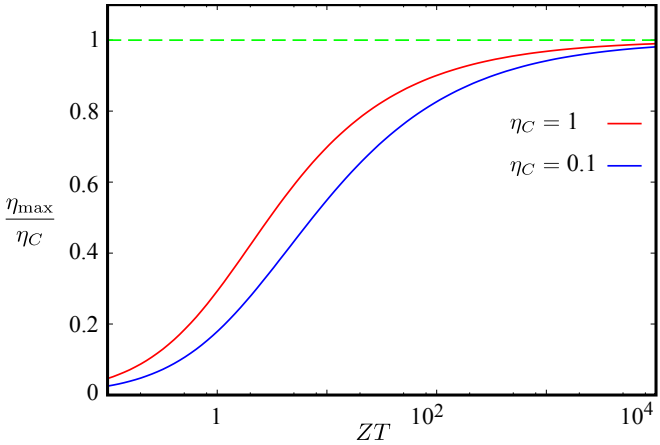

Fig. 5. The efficiency $\eta_{\text{max}}$ of a thermocouple as a function of the figure of merit $ZT$ and Carnot efficieny $\eta_C$.

in the limit of $ZT \to \infty$, while for small $ZT$, the efficiency reduces to

$$\eta_{\text{max}} \to \frac{\Delta T}{T_H} \frac{ZT}{4} \frac{1}{1-\frac{\Delta T}{2T_H}} = \frac{\Delta T}{T} \frac{ZT}{4}.$$ (28)

The figure of merit (25), depends on the device dimensions $A_p$, $A_n$, $l_p$, and $l_n$. It can be maximized in devices with $A_p l_n/A_n l_p = \sqrt{\kappa_n \sigma_n/\kappa_p \sigma_p}$, for which

$$(ZT)_{\text{max}} = \left(\frac{S_p - S_n}{\sqrt{\frac{\kappa_p}{\sigma_p}} + \sqrt{\frac{\kappa_n}{\sigma_n}}}\right)^2 T.$$ (29)

When $S_p = -S_n$ and $\kappa_p/\sigma_p = \kappa_n/\sigma_n$, the $ZT$ equals that of the individual conductors:

$$ZT_{p,n} = \frac{\sigma_{p,n} S_{p,n}^2 T}{\kappa_{p,n}}.$$ (30)

When $\kappa/\sigma$ is negligibly small for one of the conductors, e.g. when it is a superconductor [19], $ZT$ is governed solely by the other material. The devices with high $ZT$ may be engineered by tuning the material properties, e.g., by doping [24], or employing novel materials such as topological insulators [56]–[58]. In nanoscale materials such as wires and point contacts, the figure of merit may be enhanced due to large energy derivatives at the conductance steps of quantum point contacts [40] or van Hove singularities in the density of states of quantum wires [25], [26]. To find materials with $ZT \gtrsim 3$ at ambient temperatures is one of the grand challenges for condensed matter physics; no fundamental laws appear to prohibit it.

## III. SPIN CALORITRONICS

In this Section, we summarize recent developments in the field of spin caloritronics. This branch of science studies the generation and control of coupled charge, spin, and heat currents [4], which operates in the overlap region of thermoelectrics and spintronics/nanomagnetism. In Section III-A we discuss the thermoelectric properties of metallic ferromagnets in the two-current model of spin-up and spin-down electrons. In Section III-B we turn to the power conversion schemes that utilize the so-called spin Seebeck effect.

### A. Spin-dependent Seebeck and Peltier effects

In Sec. II we considered spin-degenerate conductors with thermoelectric properties that differ trivially from those of hypothetical carriers without spin. In a normal metal and in the absence of a magnetic field the states with up and down spins are equally occupied. There is no net magnetic moment and an electric field drives a pure charge current. In this section we discuss thermoelectric transport when spin degeneracy is lifted by a strong external magnetic field or, in a ferromagnet, by the exchange interaction. The density of states, mobilities, and conductances of the two spin states then differ, hence an electric field $E$ drives different currents of both spins; the current inside a conductor is spin-polarized. The charge current is the sum of spin-up and spin-down currents $J_c = \sum_{\varsigma} J_{\varsigma}$, where $J_{\varsigma} = \sigma_{\varsigma} E$, ($\varsigma = \{\uparrow, \downarrow\} = \{+, -\}$) for up and down spin with intrinsic angular momentum projection $\varsigma h/(4\pi) = \varsigma\hbar/2$, where $h$ is Planck's constant. The spin current is the difference $J_s = \sum_{\varsigma} \varsigma J_{\varsigma}$, representing the flow of angular momentum $\hbar J_s/(2e)$. The up spin electron has positive angular momentum but in vacuum and most materials negative magnetic momentum. As shown in Table I, for positive-charged carrier, $J_{\uparrow}$ ($J_{\downarrow}$) refers to the carrier current with spin parallel (antiparallel) spin quantization direction (magnetic field)., see. On the other hand for negative-charged carrier, $-J_{\uparrow}$ ($-J_{\downarrow}$) refers to the carrier current with spin antiparallel (parallel) to magnetic field. The electric field couples only to the charge, so the driving force in (the bulk of) paramagnetic metals is the same for both spins.

A difference in the voltage for the majority and minority spins $V_{\uparrow(\downarrow)}$ in metals out of equilibrium is called spin accumulation $V_s = V_{\uparrow} - V_{\downarrow}$. In the steady state, $V_s$ obeys the spin-diffusion equation:

$$\nabla^2 V_s = \frac{V_s}{\lambda^2},$$ (31)

where $\lambda$ is spin-flip diffusion length. In normal metals $\lambda = \sqrt{D\tau_{sf}}$, where $\tau_{sf}$ is the spin-flip relaxation time. The diffusion constant $D = v_F^2 \tau/3$, where $v_F$ is the Fermi velocity and $\tau$ the transport scattering time. The temperatures may be also different for up and down spins [21]–[23], defining an observable spin heat accumulation [11] that is also governed by a diffusion equation. We disregard this complication for simplicity in the following by assuming $T = T_\uparrow = T_\downarrow$.

TABLE I
Sign definition of spin and charge currents.

| Carrier's charge | positive | negative |
|---|---|---|
| Charge current $J_c$ | | |
| Spin current $J_s$ | | |
| $J_\uparrow = \frac{1}{2}\left(J_c + J_s\right)$ | | |
| $J_\downarrow = \frac{1}{2}\left(J_c - J_s\right)$ | | |

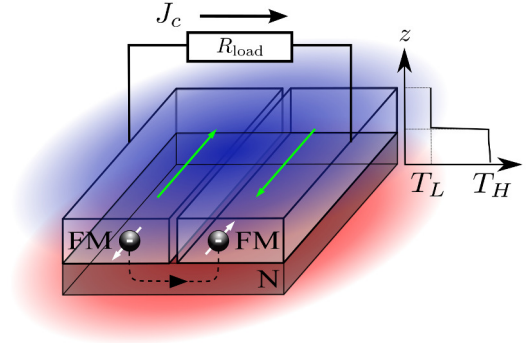

Fig. 6. A spin valve consisting of a normal metal (N) and two ferromagnetic metal (FM) contacts that are connected by a load resistance $R_{\text{load}}$. A temperature difference $\Delta T$ between N and FMs drives a charge current through the circuit.

Let us now focus on the interface between a ferromagnetic and a normal metal, denoted as FM|N. An electric field or temperature difference generates a spin accumulation in proximity of the interface that is governed by the spin-diffusion equation. We denote the drop between the spin-dependent voltage at an interface $\Delta V_\varsigma$. Since we assume $\Delta T$, the temperature drop over the interface, to not depend on spin, only the total heat current $J_Q$ is relevant. Analogous to (3) and (4) we then arrive at spin-dependent thermoelectric relations

$$J_\varsigma = -G_\varsigma\left(\Delta V_\varsigma + S_\varsigma \Delta T\right), \tag{32}$$

$$J_Q = -K\Delta T + T\sum_\kappa S_\varsigma J_\varsigma. \tag{33}$$

Here $G_\varsigma$ are the spin-dependent electric interface conductances. The spin-dependent interface Seebeck coefficients $S_\varsigma$ read to lowest order in the Sommerfeld expansion and analogous with (9)

$$S_\varsigma = -eL_0 T \frac{\partial_\varepsilon G_\varsigma\left(\varepsilon\right)|_{\varepsilon_F}}{G_\varsigma(\varepsilon_F)}, \tag{34}$$

Eqs. (32) and (33) can be rewritten as a matrix that satisfies Onsager symmetry [22], [34]:

$$\begin{pmatrix} J_c \\ J_s \\ J_Q \end{pmatrix} = G \begin{pmatrix} 1 & P & ST \\ P & 1 & P'ST \\ ST & P'ST & \mathcal{K}T/G \end{pmatrix} \begin{pmatrix} -\Delta V \\ -\Delta V_s/2 \\ -\Delta T/T \end{pmatrix}. \tag{35}$$

Here $V = (V_\uparrow + V_\downarrow)/2$ is the voltage; $\Delta V$, $\Delta V_s$ and $\Delta T$ denote $V^{\text{N}} - V^{\text{FM}}$, $V_s^{\text{N}} - V_s^{\text{FM}}$ and $T_{\text{N}} - T_{\text{FM}}$, respectively; $\mathcal{K}$ is the heat conductance for $\Delta V_{\uparrow(\downarrow)} = 0$. The total electric conductance is $G = G_\uparrow + G_\downarrow$, $S$ is the total interface Seebeck coefficient, $P$ is the spin polarization, and $P'$ is the polarization of the product $GS$:

$$S = \frac{G_\uparrow S_\uparrow + G_\downarrow S_\downarrow}{G_\uparrow + G_\downarrow}, \tag{36}$$

$$P = \frac{G_\uparrow - G_\downarrow}{G_\uparrow + G_\downarrow}, \tag{37}$$

$$P' = \frac{G_\uparrow S_\uparrow - G_\downarrow S_\downarrow}{G_\uparrow S_\uparrow + G_\downarrow S_\downarrow}. \tag{38}$$

At the interfaces with so-called half-metals $P = P' = 1$, $G_\downarrow = 0$, and $G_\uparrow = G$. In general case $P'$ can have any (real) value since the denominator may vanish. In the Sommerfeld approximation $P'$ takes the form

$$P' = \frac{\partial_\varepsilon G_\uparrow - \partial_\varepsilon G_\downarrow}{\partial_\varepsilon G_\uparrow + \partial_\varepsilon G_\downarrow}\bigg|_{\varepsilon_F}. \tag{39}$$

Hu et al. [29] report an enhanced $P'$ for CoFeAl since $\partial_\varepsilon G_\uparrow$ and $\partial_\varepsilon G_\downarrow$ have opposite sign. The heat conductance is also affected by the spin polarization:

$$\mathcal{K} = -\left[\frac{J_Q}{\Delta T}\right]_{\Delta V = \Delta V_s = 0} = K + \frac{1 + P'^2 - 2P'P}{1 - P^2}GS^2T, \tag{40}$$

with

$$K = -\left[\frac{J_Q}{\Delta T}\right]_{J_c = J_s = 0} \tag{41}$$

In the limit $P, P' \to 0$ we recover $ZT$ for conventional thermoelectrics (25), i.e.

$$ZT_{P=P'=0} = \frac{GS^2T}{\mathcal{K}_{P=P'=0} - GS^2T} = \frac{GS^2T}{K}. \tag{42}$$

The figure of merit of heat engines susceptible to more than two forces can be maximized by a proper linear combination of the input variables. In the case of spin-dependent thermoelectrics, a better Peltier cooling can be achieved by simultaneously applying a voltage together with a spin accumulation [33].

Isolated FM|N interfaces can be studied in principle in heterostructure point contacts. However, usually they are part of a larger device such as a spin valve or multilayer. The voltage and temperature distributions in disordered metals are governed by diffusion equations with solutions that have to be knitted together at the interface. The above relations provide the necessary boundary conditions for solving the diffusion equations. Interfaces also play a role in devices such as spin valve structures as shown in Fig. 6. The spin, charge, and heat transport properties depend on the relative angle between the two magnetizations [22]. Let us consider a lateral spin valve with cold ferromagnetic contacts on top of a hot normal metal layer (a similar configuration will be considered for the spin Seebeck generator in the next section). For illustrative

reasons the two ferromagnets and their contacts are assumed the same, except for Seebeck coefficients that have opposite sign $S_2 = -S_1$ (for $S_2 = S_1$ the Seebeck effect vanishes). We also assume that the temperature drop is dominated by the interface because of a large Kapitza resistance. Below we discuss the figure of merit for limiting cases of parallel and antiparallel magnetizations in the absence of spin-flip scattering.

For $\lambda \to \infty$ and in the parallel configuration of a symmetric spin valve the spin current is maximal, while the spin accumulation vanishes. The situation is opposite for the antiparallel configuration, with zero spin current but maximum spin accumulation. The thermoelectric response matrix (35) for spin valves with parallel configuration ($P_1 = P_2 = P$ and $P'_1 = P'_2 = P'$) then reads

$$\begin{pmatrix} J_c \\ J_Q \end{pmatrix} = \begin{pmatrix} G & GS \\ GST & \mathcal{K} \end{pmatrix} \begin{pmatrix} -\Delta V \\ -\Delta T \end{pmatrix}, \quad (43)$$

where $G = G_1 G_2/(G_1 + G_2)$, $S = S_1 - S_2$, and $K = K_1 + K_2$. Results for the antiparallel configuration are obtained by setting $P_1 = -P_2 = P$ and $P'_1 = -P'_2 = P'$:

$$\begin{pmatrix} J_c \\ J_Q \end{pmatrix} = \begin{pmatrix} (1-P^2)G/2 & (1-PP')GS \\ (1-PP')GST & K + \frac{(1-PP')^2}{1-P^2}GS^2 T \end{pmatrix} \begin{pmatrix} -\Delta V \\ -\Delta T \end{pmatrix}, \quad (44)$$

The figures of merit become

$$ZT_{\mathrm{par}} = \frac{GS^2 T}{K + \frac{(P-P')^2}{1-P^2} GS^2 T} < \frac{GS^2 T}{K}, \quad (45)$$

$$ZT_{\mathrm{apar}} = \frac{(1-PP')^2}{1-P^2} \frac{GS^2 T}{K}. \quad (46)$$

The parallel configuration is less efficient than that of a normal metal trilayer with equivalent interface conductances and thermopowers. On the other hand the antiparallel spin valve may actually be a more efficient power generator, as shown in Fig. 7.

An antiparallel spin valve with material (combination) that satisfies

$$P > \frac{2P'}{1+P'^2} \quad (47)$$

would lead to enhanced figure of merit, *e.g.* for conservative $P = 0.8$ and $P' = -0.8$, $ZT \approx 4.5 \times GS^2T/K$, illustrating that manipulating spin currents may improve thermoelectric efficiencies.

### B. Spin Seebeck and Peltier effects

In Section III-A, we discussed the spin-dependent Seebeck effect associated with charge currents. In this section, we turn to pure spin current generated by temperature differences in bilayers of a magnetic insulator (FI) and a normal metal [60] that is detected in the normal metal N by the inverse spin Hall effect (ISHE) in terms of an emf transverse to

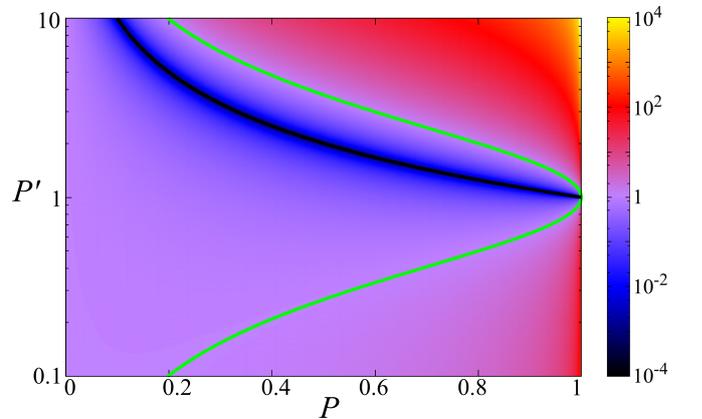

Figure 7. Figure of merit $ZT/(GS^2T/K)$ of an antiparallel spin valve (as in Fig. 6, assumed to be symmetric except for interface Seebeck coefficients that have opposite sign) as a function of the spin polarizations $P = (G^\uparrow - G^\downarrow)/(G^\uparrow + G^\downarrow)$ and $P' = (G^\uparrow S^\uparrow - G^\downarrow S^\downarrow)/(G^\uparrow S^\uparrow + G^\downarrow S^\downarrow)$. The black and green lines represent $ZT = 0$ and $ZT = GS^2T/K$, respectively. Materials with $(P, P')$ values to the right of the green line lead to improved power conversion.

the spin current direction and spin polarization. The SSE has been observed in two principal configurations, i.e. "transverse" (Fig. 8) and "longitudinal" (Fig. 8), referring to the relative direction of temperature gradient and spin current. The SSE has been discovered in the "transverse" configuration in permalloy [61], ferromagnetic semiconductors (GaMnAs) [31], electrically insulating yttrium iron garnet (YIG) [62], and

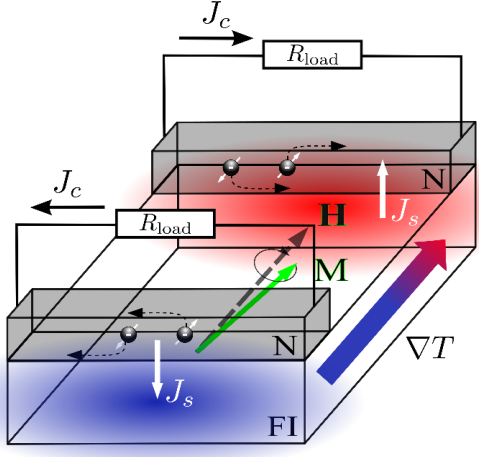

Fig. 8. The transverse configuration for the spin Seebeck effect. Spin currents $J_s$ are generated at the interfaces between the ferromagnet (FI) and metal (N) on the hot and the cold sides, which are found to be of opposite sign [60], [61]. Spin current directions are normal to the temperature gradient $\boldsymbol{\nabla} T$. The spin currents in N are efficiently converted to Hall voltages by the inverse spin Hall effect (ISHE) when $\mathbf{M} \| \boldsymbol{\nabla} T$.

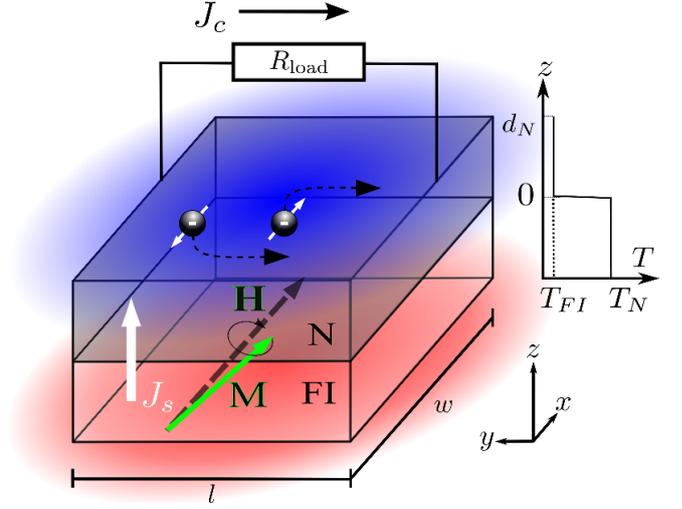

Fig. 9. The longitudinal configuration for the spin Seebeck effect, in which spin current and temperature gradient are parallel. A finite FI|N interface heat conductance $G_K$ generates a temperature drop at the interface. $J_s$ is converted into a transverse voltage or charge current $J_c$ by the ISHE.

Heusler alloys [5]. It is caused by a net spin current caused by a difference between the temperature of the order parameter in the ferromagnet ("magnon temperature") and the electrons in the normal metal contact [68].

The transverse SSE is still not well understood [3], [43], [50], presumably because of a sensitive dependence on the details of the temperature distribution as well as the role non-thermalized phonons in the substrate [1], [55]. We focus here on the longitudinal configuration (Fig. 9) in which the spin and heat currents are normal to the interfaces [60], which has been reproduced by many groups. When the ferromagnet is a metal, the SSE occurs in parallel with the anomalous Nernst effect. This is not an issue for a ferromagnetic insulator such as YIG, but from a device point of view SSE and anomalous Nernst effect may both be useful [38]. The longitudinal SSE as detected by the ISHE converts thermal energy into electric power that is proportional to the sample area and therefore does not require complicated thermopile structuring. It is attractive for applications in large-area thermoelectric power generators that are made by simply coating suitable substrates [35], [47].

We defined the spin current earlier as the difference of spin up and down electric current. Generally speaking, the spin current is a tensor spanned by its direction and polarization. Here we define a spin current by the direction vector $\mathbf{J}_s$ while its polarization is denote by $\boldsymbol{\sigma}$. When a spin current from N encounters a magnetic insulator polarized parallel to the magnetization, it is fully reflected. However, a spin current can flow through an FI|N interface mediated by an interface exchange interaction. It is absorbed at the interface, thereby exerting a torque on the magnetization that can excite a spin current in the ferromagnet in the form of spin waves. Vice versa, a spin current from the ferromagnet can be injected into the normal metal via spin pumping [49]. The SSE is caused by an imbalance between the spin current $J_{sp}$ pumped from the ferromagnet into the normal metal by magnetic thermal noise and the spin current $J_{fl}$, which is the component of thermal (Johnson-Nyquist) spin current noise with polarization normal to the magnetization. These two currents cancel at thermal equilibrium [17], [67], but not in the presence of a temperature difference over the interface.

The spin current pumped out of the ferromagnet into a normal metal reads [17], [67]

$$e J_{sp} \boldsymbol{\sigma} = \frac{\hbar}{4\pi} \langle G_r \mathbf{m} \times \dot{\mathbf{m}} + G_i \dot{\mathbf{m}} \rangle ,\qquad (48)$$

where $\mathbf{m}$ is the magnetization direction at the interface, $\langle \cdots \rangle$ is a time or ensemble average, and $G_r/G_i$ is the real/imaginary part of the spin mixing conductance, which determines the magnitude of the pumped spin current [6] in units of $\Omega^{-1}$. The thermal spin fluctuations in the metal layer generate a stochastic spin-transfer torque on the magnetization that can be expressed by a random magnetic field $\mathbf{h}'$ acting on the interface magnetization

$$e J_{fl} \boldsymbol{\sigma} = -M_s V_c \langle \mathbf{m} \times \mathbf{h}' \rangle .\qquad (49)$$

where $M_s$ is the saturation magnetization and $V_c \sim \lambda_T^3$ is a magnetic coherence volume [69] ($\lambda_T$ the de Broglie magnon thermal wave length [27]) that depends on the spin wave stiffness and weakly on temperature. The polarization of both (thermally averaged) spin currents are collinear with the equilibrium magnetization $\boldsymbol{\sigma} \| \langle \mathbf{m} \rangle$ [52] .

In thermal equilibrium $J_s = J_{sp} + J_{fl} = 0$ reflects the second law of thermodynamics and can be interpreted as a fluctuation dissipation theorem [67]. $J_{sp}$ is proportional to the magnon temperature $T_{FI}^m$, while $J_{fl}$ is proportional to the electron temperature $T_N^e$ in N. A difference between $T_{FI}^m$ and $T_N^e$, destroys the balance and a net spin current proportional to $T_{FI}^m - T_N^e$ flows through the interface. The spin current

pumped from FI to N is

$$J_s = L_S \left(T_{FI}^m - T_N^e\right), \tag{50}$$

where

$$L_S \simeq \frac{\gamma \hbar G_r k_B}{\pi e M_s} \left(\frac{k_B T_F}{4\pi D}\right)^{3/2} \tag{51}$$

is the interfacial spin Seebeck coefficient, $\gamma$ the gyromagnetic ratio, and $D$ the spin wave stiffness of the magnon dispersion $\hbar\omega = \hbar\omega_0 + Dk^2$. Adachi *et al.* [1], Hoffman *et al.* [27], and Rezende *et al.* [45] derived similar expressions by different methods. The spin mixing conductance of an FI|N interface is of the same order of magnitude as that of metallic junctions FM|N [7], [32], [66]; for YIG|Pt $G_r/A = 5 \times 10^{15}$ $\Omega^{-1}\mathrm{m}^{-2}$ (or $10^{19}$ m$^{-2}$ when spin current is represented in energy units by the conversion factor $2e^2/\hbar$).

Part of the injected spin current is reflected back into the ferromagnet by a second interface or defects, thereby generating a spin accumulation $V_s \langle\mathbf{m}\rangle$ at the interface on the N side. The current injected into FI by a spin accumulation in F reads [70]:

$$J_{bf}\langle\mathbf{m}\rangle = \langle G_r\mathbf{m} \times (\mathbf{m} \times V_s\langle\mathbf{m}\rangle) + G_i\mathbf{m} \times V_s\langle\mathbf{m}\rangle\rangle, \tag{52}$$

$$J_{bf} = -G_r(1 - \langle m_x^2\rangle)V_s = -G_S V_s/2, \tag{53}$$

where we introduced the temperature-dependent interface spin conductance [70]

$$G_S \simeq 1.3\frac{G_r}{\pi}\frac{\gamma\hbar}{M_s}\left(\frac{k_B T_F}{4\pi D}\right)^{3/2}. \tag{54}$$

Hence, the total injected spin current is

$$J_s = L_S \left(T_{FI}^m - T_N^e\right) - G_S V_s/2, \tag{55}$$

which may be compared with (50).

The temperature drop and spin accumulation at the interface that govern the thermal spin current (55) cannot be directly controlled. The interface is embedded into a larger structure or device. Heat sources at a distance generate heat currents and temperature profiles whose modelling is far from trivial. Sanders and Walton [48] introduced a two reservoir diffusion theory for coupled heat transport in the magnon-phonon system through magnetic insulators. The neglect of interface heat (Kapitza) resistances $R_\kappa$ [42] and vanishing spin and heat current at the boundaries has been relaxed in Refs. [51], [68]. However, finite $R_\kappa$ could change the qualitative interpretation of temperature profile, even in phonon-dominated transport across magnetic interface [72]. Therefore, interface heat resistance should be taken into account.

The situation is simplified when heat transport is limited by a sufficiently large $R_\kappa$. The magnon and phonons on the insulator side of the interface then have the time to equilibrate to the same temperature [2]:

$$\lim_{R_\kappa \to \infty} T_{FI}^p(0^-) = T_{FI}^m(0^-) = T_{FI}(0^-) \tag{56}$$

$$= T_N^e(0^+) + R_\kappa J_Q, \tag{57}$$

which is valid for

$$R_\kappa \gg \frac{\lambda_m}{\kappa_p A}\tanh\frac{d_{FI}}{2\lambda_m}, \tag{58}$$

where $J_Q$ is the heat current from FI to N, $d_{FI}$ is the thickness of the magnetic layer, $\kappa_p$ is the phonon heat conductivity, and $\lambda_m$ is the magnon relaxation length in the magnet. Then the spin current reads:

$$J_s(0) = L_S \left(T_{FI} - T_N\right) - G_S V_s/2e, \tag{59}$$

where $T_N \equiv T_N^e$ here and below.

The matrix of linear response coefficients for spin and heat current is completed by the spin Peltier effect with coefficient defined analogous to the conventional Peltier coefficient as

$$\Pi_S = \left[\frac{J_Q}{J_s}\right]_{\Delta T = 0} \tag{60}$$

where here $\Delta T = T_N - T_{FI}$, leading to the total heat current

$$J_Q = -G_K\Delta T + \Pi_S J_s. \tag{61}$$

We can now write a spin caloritronic linear-response relation for spin current $J_s$ and heat current $J_Q$ in a form similar to the conventional thermoelectric relations (19)

$$\begin{pmatrix} J_s \\ J_Q \end{pmatrix} = G_S \begin{pmatrix} 1 & S_S \\ \Pi_S & G_K/G_S + S_S^2 T \end{pmatrix} \begin{pmatrix} -V_s/2 \\ T_{FI} - T_N \end{pmatrix}, \tag{62}$$

where $S_S = -(V_s/(2\Delta T))_{J_s=0} = L_S/G_S$ is the spin Seebeck coefficient that (for parabolic spin wave dispersions) has the universal value $S_S \approx 0.77\ k_B/e = 65\ \mu\mathrm{V K}^{-1}$ [70], and $G_K = -(J_Q/\Delta T)_{J_s=0}$ is the Kapitza conductance, which is the inverse of the Kapitza resistance $G_K = 1/R_K$. $\Pi_S$ is the spin Peltier coefficient that according to the Onsager symmetry for insulator spin caloritronics is related to the spin Seebeck coefficient $\Pi_S = S_S T$. The spin Peltier effect observed by Flipse *et al.* [16] appears to obey this relation, although it has not yet been demonstrated to hold for one and the same sample.

The pumped spin current can be converted into electric energy by two methods. By utilizing the ISHE [59] or by using the spin valve that is based on the spin filtering mechanism [8]. In Sections III-C and III-D we discuss the spin-charge conversion mechanisms. The corresponding figures of merit of are derived in Section IV.

### C. Spin charge conversion by inverse spin Hall effect

The spin Hall effect refers to the generation of a spin current transverse to an applied electric current due to the spin-orbit interaction. The polarization of the generated spin current is normal to both spin and charge current directions [28]. The inverse spin Hall effect converts a spin current into an electric current (Fig. 10a). Hence, it can be utilized to generate electric power from the SSE spin current. We can, reciprocally, inject a spin Hall spin current into the FI by driving a current through the normal metal film. The spin Peltier effect then cools or heats the sample, depending on the relative orientation of current and magnetization directions.

In a FI|N bilayer the polarization of the pumped spin current is parallel to the magnetization $\mathbf{m}$ (see Fig. 9). In a closed circuit, current-biased configuration in which $\nabla V = 0$, the

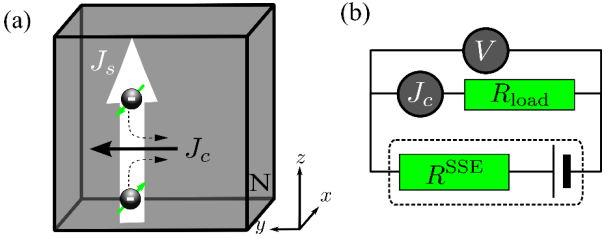

Fig. 10. (a) Inverse spin Hall effect: the conversion of spin currents into transverse charge currents by the spin-orbit interaction in a normal metal (N). (b) Equivalent circuit of spin Seebeck power conversion: the emf generated by the inverse spin Hall effect system is connected to a load resistance $R_{\mathrm{load}}$ is equivalent to a battery with voltage $S_{\mathrm{ISHE}}^{\mathrm{SSE}} \Delta T$ and internal resistance $R_{\mathrm{ISHE}}^{\mathrm{SSE}}$.

spin current density is fully converted into a charge current density $\mathbf{j}_c$ according to

$$\mathbf{j}_c = \theta_{\mathrm{SH}} \mathbf{j}_s \times \langle \mathbf{m} \rangle , \qquad (63)$$

where $\mathbf{j}_s$ is the spin current density (direction), the spin current polarization is parallel to the FI's magnetization $\langle \mathbf{m} \rangle$, and $\theta_{\mathrm{SH}}$ is the spin Hall angle [54]. The charge current induces an electric field $\nabla V$, leading to an Onsager symmetric charge-spin current relation in N that illustrates the Onsager reciprocity between the spin Hall effect and its inverse [9], [54]

$$\begin{pmatrix} \mathbf{j}_c \\ \mathbf{j}_s \end{pmatrix} = \sigma_N \begin{pmatrix} -1 & \theta_{\mathrm{SH}} \langle \mathbf{m} \rangle \times \\ \theta_{\mathrm{SH}} \langle \mathbf{m} \rangle \times & -1 \end{pmatrix} \begin{pmatrix} \nabla V \\ \nabla V_s/2 \end{pmatrix} , \qquad (64)$$

where $\sigma_N$ is the electric conductivity.

In Fig. 9, $\langle \mathbf{m} \rangle$ and therefore the polarization of the SSE spin current are along $\mathbf{x}$. The spin angular momentum flows in the $z$-direction: $\mathbf{j}_s = j_s \mathbf{z}$ and $\nabla V_s = (\partial_z V_s) \mathbf{z}$. Therefore, with $\mathbf{j}_c = j_c \mathbf{y}$ and $\nabla V = (\partial_y V) \mathbf{y}$, Eq. (64) simplifies to

$$\begin{pmatrix} j_c \\ j_s \end{pmatrix} = \sigma_N \begin{pmatrix} 1 & -\theta_{\mathrm{SH}} \\ \theta_{\mathrm{SH}} & 1 \end{pmatrix} \begin{pmatrix} -\partial_y V \\ -\partial_z V_s/2 \end{pmatrix} . \qquad (65)$$

The anti-symmetry of the response matrix in Eq. (65) will be discussed in Section IV-A. The spatial dependence of the spin accumulation is governed by the spin-flip diffusion equation (31) with the following boundary conditions: At $z = 0$ a spin current is injected by the SSE and reduced by the backflow, whereas it vanishes at the interface to the vacuum ($z = d_N$)

$$J_s(0) = G_S S_S (T_{FI} - T_N) - G_S V_s(0)/2, \qquad (66)$$

$$J_s(d_N) = 0. \qquad (67)$$

The solutions

$$\frac{V_s}{2e} = \frac{\theta_{\mathrm{SH}} V \lambda}{l} \frac{G_N \cosh \frac{z}{\lambda} + G_S \sinh \frac{z}{\lambda}}{G_N \sinh \frac{d_N}{\lambda} + G_S \cosh \frac{d_N}{\lambda}}$$
$$- \frac{G_S S_S \Delta T + \theta_{\mathrm{SH}} G_N V \lambda/l}{G_N \sinh \frac{d_N}{\lambda} + G_S \cosh \frac{d_N}{\lambda}} \cosh \frac{z - d_N}{\lambda} \qquad (68)$$

depend on the conductance of the "magnetically active" slice $G_N = \sigma_N w l / \lambda$, where $w$ and $l$ are the width and length of the normal metal film, see Fig. 9. For constant transverse electric field $\partial_y V$ the total ISHE voltage amounts to $\Delta V = -l \partial_y V$.

The spin current $J_s = w l j_s$ and the integrated transverse charge current $J_c = -w \int j_c(z) dz$ in N then read

$$J_s(z) = G_N \left[ \frac{\theta_{\mathrm{SH}} G_N V \lambda/l + G_S S_S \Delta T}{G_N \sinh \frac{d_N}{\lambda} + G_S \cosh \frac{d_N}{\lambda}} \sinh \frac{z - d_N}{\lambda} \right.$$
$$\left. + \frac{\lambda \theta_{\mathrm{SH}} V}{l} \left( 1 - \frac{G_S \cosh \frac{z}{\lambda} + G_N \sinh \frac{z}{\lambda}}{G_N \sinh \frac{d_N}{\lambda} + G_S \cosh \frac{d_N}{\lambda}} \right) \right] , \quad (69)$$

$$J_c = \frac{G_N \lambda}{l} \left[ \frac{\theta_{\mathrm{SH}} G_S S_S (T_{FI} - T_N) \tanh \frac{d_N}{2\lambda}}{G_N + G_S \coth \frac{d_N}{\lambda}} \right.$$
$$\left. - \frac{d_N V}{l} - \frac{\theta_{\mathrm{SH}}^2 \lambda V}{l} \frac{G_S + 2 G_N \tanh \frac{d_N}{2\lambda}}{G_N + G_S \coth \frac{d_N}{\lambda}} \right] . \qquad (70)$$

A heat-harvesting device is basically a battery with a maximum voltage $S_{\mathrm{ISHE}}^{\mathrm{SSE}} |\Delta T|$ and internal resistance $R_{\mathrm{ISHE}}^{\mathrm{SSE}}$ (Fig. 10). The output voltage $V$ depends on the load resistance $R_{\mathrm{load}}$

$$\Delta V = \frac{R_{\mathrm{load}}}{R_{\mathrm{load}} + R_{\mathrm{ISHE}}^{\mathrm{SSE}}} S_{\mathrm{ISHE}}^{\mathrm{SSE}} (T_{FI} - T_N) , \qquad (71)$$

$$\frac{1}{R_{\mathrm{ISHE}}^{\mathrm{SSE}}} = \frac{\sigma_N w d_N}{l} \left( 1 + \frac{\theta_{\mathrm{SH}}^2 \lambda}{d_N} \frac{G_S + 2 G_N \tanh \frac{d_N}{2\lambda}}{G_N + G_S \coth \frac{d_N}{\lambda}} \right) , \qquad (72)$$

$$S_{\mathrm{ISHE}}^{\mathrm{SSE}} = R_{\mathrm{ISHE}}^{\mathrm{SSE}} \frac{\theta_{\mathrm{SH}} \sigma_N w G_S S_S \tanh \frac{d_N}{2\lambda}}{G_N + G_S \coth \frac{d_N}{\lambda}} \propto l. \qquad (73)$$

When $R_{\mathrm{load}} \to \infty$, $\Delta V \to S_{\mathrm{ISHE}}^{\mathrm{SSE}} \Delta T \propto l$: the output voltage can be increased by elongating the device in the direction perpendicular to both the temperature gradient and $\langle \mathbf{m} \rangle$. In the limit of a thick N layer, $S_{\mathrm{ISHE}}^{\mathrm{SSE}} \Delta T$ vanishes as $1/d_N$ since it is short-circuited by the magnetically inactive part of N. It also vanishes in the opposite limit of very thin N films, $d_N \ll \lambda$, in which there is not sufficient volume to convert the spin current into charge current by means of ISHE.

The maximum electric power transferred to the load resistance $R_{\mathrm{load}}$ is proportional to the FI|N contact area:

$$W = \left( \frac{S_{\mathrm{ISHE}}^{\mathrm{SSE}} \Delta T}{R_{\mathrm{load}} + R_{\mathrm{ISHE}}^{\mathrm{SSE}}} \right)^2 R_{\mathrm{load}} \propto wl. \qquad (74)$$

$W$ is proportional to the interface area, which implies that a larger power can be generated simply by scaling-up the device.

### D. Detection by spin valve

The spin-motive force generated by the temperature difference over the interface in N can also be harvested by ferromagnetic metal contacts [8]. The physics here is similar to the giant magnetoresistance (GMR) [15]. For best results the spacer layer should then be chosen from metals with weak spin-orbit interaction and long spin-flip diffusion lengths such as copper, since the spin Hall effect is not required. A large spin mixing conductance for the Cu|YIG interface has been predicted by first principles calculations [32] and found experimentally for Au|YIG [7]. Experiments on non-local spin valves based on Cu on YIG suggested lower values, presumably due to interface contamination [63]. Similar experiments

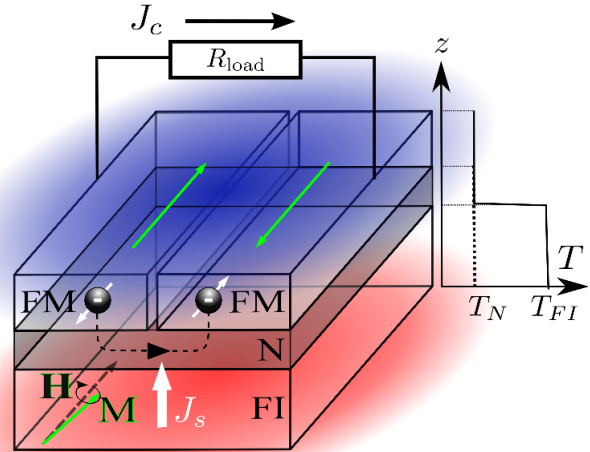

Fig. 11. A schematic view of the lateral spin-valve based spin Seebeck power conversion device. Two FM contacts with antiparallel magnetizations are attached to the FI|N bilayer device from Fig. 9. The spin current pumped from FI to N generates a spin accumulation that drives a charge current through the spin valve, thereby generating a voltage that is discharged over the load resistance. Since the ISHE is not required, an N with long spin-flip diffusion length is beneficial for the device performance.

found large spin mixing conductance for the Al|YIG interface [12].

As indicated in Fig. 11, the spin current is now injected into the spacer layer of a lateral metallic spin valve with ferromagnetic contacts in an antiparallel configuration collinear to the FI magnetization. The equivalent circuit is shown in Fig. 12. A spin accumulation injected into the spacer by the spin Seebeck effect then generates a charge current through the spin valve. We can operate this device as a spin Peltier cooler by externally driving a current through the spin valve.

In the structure sketched in Fig. 11 two antiparallel FM layers contact the N side of the FI|N bilayer. We disregard spin-flip in N, which is allowed as long as the electron dwell time is short compare to the spin-flip relaxation time. We assume that the conductivity of the normal metal N (such as Cu) is much higher than that of the metallic ferromagnet (such as permalloy). The spin-dependent voltage $V_\uparrow$ and $V_\downarrow$ are then constant, leading to the equivalent circuit in Fig. 12. The N|FM interface areas are taken to be $A$ and that of the FI|N interface approximately $2A$. The N|FM interface conductance $G_I = G_{I\uparrow} + G_{I\downarrow}$ includes the magnetically active thickness of the FM and is the same for both contacts.

Assuming that the currents only flow in the $z$ direction (normal to the interfaces) and that the temperature variation in N and FM is very small compared to the temperature drop at the FI|N interface (see Fig. 11), the spin-charge current relations at the N|FM interface can be written as

$$\begin{pmatrix} J_c^\varkappa \\ J_s^\varkappa \end{pmatrix} = G_I \begin{pmatrix} 1 & \varkappa P \\ \varkappa P & 1 \end{pmatrix} \begin{pmatrix} V^\varkappa - V^N \\ (V_s^\varkappa - V_s^N)/2 \end{pmatrix}, \quad (75)$$

where $\varkappa = \pm$ indicates the two FMs with magnetizations parallel or antiparallel to the FI, and $P$ is the FM conductance polarization. The response matrix is symmetric here, unlike to

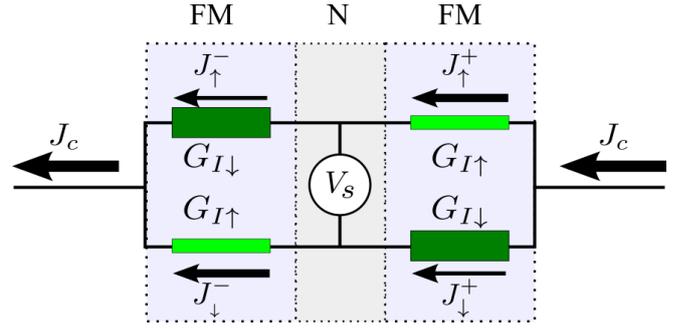

Fig. 12. Equivalent circuit for the spin-valve generate from Fig. 11. The thermally induced spin accumulation $V_s$ in N drives a charge current $J_c$.

Eq. (65). Assuming zero spin accumulation in the FM contacts, the charge currents through each contact read

$$J_c^\varkappa = -\varkappa J_c = G_I \left( V^\varkappa - V^N \right) - \varkappa P G_I \frac{V_s^N}{2} \quad (76)$$

by charge conservation. It follows from (75) that the FM's generate voltage differences $V^\varkappa = \left( V^\varkappa - V^N \right)/e = -V^{-\varkappa}$ that are equal in magnitude but of opposite sign. The total output voltage is then $\Delta V = V^+ - V^-$.

Since we assume conservation of spin in N, the total spin current out of FI, $J_s^{FI}$ in Eq. (62), and the FM's, $J_s^\varkappa = -G_I(\Delta V + V_s^N)/2$, should vanish. The spin accumulation in N can be then expressed in terms of $\Delta T$ and $\Delta V$ as:

$$\frac{V_s^N}{2} = \frac{PG_I \Delta V - G_S S_S \Delta T}{G_S + 2G_I}. \quad (77)$$

By substituting (68) into (75)

$$J_c = -\frac{PG_I G_S S_S \Delta T}{G_S + 2G_I} - \frac{G_I \Delta V}{2} \frac{G_S + 2(1 - P^2)G_I}{G_S + 2G_I}. \quad (78)$$

The effective electric circuit for the spin-valve based power conversion is still as in Fig. 10b, but with parameters:

$$\Delta V = \frac{R_{\text{load}}}{R_{\text{load}} + R_{\text{SV}}^{\text{SSE}}} S_{\text{SV}}^{\text{SSE}} (T_{FI} - T_N), \quad (79)$$

$$R_{\text{SV}}^{\text{SSE}} = \frac{2}{G_I} \frac{G_S + 2G_I}{G_S + 2(1 - P^2)G_I}, \quad (80)$$

$$S_{\text{SV}}^{\text{SSE}} = \frac{2PG_S S_S}{G_S + 2(1 - P^2)G_I}. \quad (81)$$

Efficiency $\eta = W/J_Q = \left( \Delta V/(R_{\text{load}} + R_{\text{SV}}^{\text{SSE}}) \right)^2 R_{\text{load}}/J_Q$ is sample-scale invariant, which allows miniaturizing the device to the nanoscale regime without degradation.

## IV. SPIN SEEBECK FIGURE OF MERIT

We now derive the spin Seebeck/Peltier figure of merit from the spin caloritronic response relations (62) for the two spin-charge conversion methods. We refrain from defining a figure of merit for the spin current/accumulation generation, which is straightforward but not very useful. Both SSE power-conversion schemes can be represented by the same effective electric circuit (Fig 10). We still have to formulate the heat current $J_Q(V_s, \Delta T)$ in (62) as a function of $\Delta T = T_N - T_{FI}$ and $V_s$.

The relation between currents and forces in our devices reads quite generally

$$
\begin{pmatrix} J_c \\ J_Q \end{pmatrix} = G^{\mathrm{SSE}} \begin{pmatrix} 1 & S^{\mathrm{SSE}} \\ \Pi^{\mathrm{SSE}} & \frac{K^{\mathrm{SSE}}}{G^{\mathrm{SSE}}} + \Pi^{\mathrm{SSE}} S^{\mathrm{SSE}} \end{pmatrix} \begin{pmatrix} -\Delta V \\ -\Delta T \end{pmatrix}
\tag{82}
$$

Here $K^{SSE} = J_Q / (T_{FI} - T_N)|_{J_c=0}$ is the effective heat conductance for an open electric circuit and $\Pi^{\mathrm{SSE}} = (J_Q/J_c)_{\Delta T=0}$ is the (charge) Peltier coefficient. (82) is analogous to (19) for spin-dependent thermoelectrics. From (62) and (82)

$$
\Pi^{\mathrm{SSE}} = G_S S_S T R^{\mathrm{SSE}} \left[ \frac{V_s^N(0)}{2\Delta V} \right]_{\Delta T=0},
\tag{83}
$$

$$
K^{\mathrm{SSE}} = G_K + G_S S_S^2 T + G_S S_S T \left[ \frac{V_s^N(0)}{2\Delta T} \right]_{\Delta V = S^{\mathrm{SSE}}\Delta T}.
\tag{84}
$$

For $\Delta T > 0$ the heat current $J_Q < 0$. The conversion efficiency $\eta^{\mathrm{SSE}} = W/J_Q^{\mathrm{IN}}$ is shown below to be a function of $J_c$. By optimizing $J_c$, the maximum efficiency of our spin caloritronics power generators turns out to have the same form as that of the conventional thermoelectric generator (24):

$$
\eta_{max}^{\mathrm{SSE}} = \eta_C \frac{\sqrt{1+(ZT)^{\mathrm{SSE}}} - 1}{\sqrt{1+(ZT)^{\mathrm{SSE}}} + 1 - \eta_C}.
\tag{85}
$$

The value of the spin Seebeck figure of merit $(ZT)^{\mathrm{SSE}}$ for each spin-charge conversion scheme is discussed in the following sections.

### A. Detection by inverse spin Hall effect

The spin Seebeck/Peltier figure of merit of the bilayer from Section III-C requires knowledge of the effective heat conductance $K_{\mathrm{ISHE}}^{\mathrm{SSE}}$ and Peltier coefficient $\Pi_{\mathrm{ISHE}}^{\mathrm{SSE}}$. By substituting (68) into (83) and (84), we arrive at

$$
\Pi_{\mathrm{ISHE}}^{\mathrm{SSE}} = -S_{\mathrm{ISHE}}^{\mathrm{SSE}} T
\tag{86}
$$

$$
K_{\mathrm{ISHE}}^{\mathrm{SSE}} = G_K + \frac{G_N G_S S_S^2 T}{G_N + G_S \coth \frac{d_N}{\lambda_N}} + \frac{\left(S_{\mathrm{ISHE}}^{\mathrm{SSE}}\right)^2 T}{R_{\mathrm{ISHE}}^{\mathrm{SSE}}}.
\tag{87}
$$

The different sign of $\Pi_{\mathrm{ISHE}}^{\mathrm{SSE}}$ and $S_{\mathrm{ISHE}}^{\mathrm{SSE}}$ is caused by the antisymmetric components in (65). Negative sign means that if a charge current inputted in the same direction as the output of Seebeck effect, the output Peltier heat current should be in the opposite direction in respect to the direction of input heat current of Seebeck effect. The power conversion efficiency can then be represented as a function of $J_c$,

$$
\eta(J_c) = \frac{J_c S_{\mathrm{ISHE}}^{\mathrm{SSE}} \Delta T - J_c^2 R_{\mathrm{ISHE}}^{\mathrm{SSE}}}{K\Delta T - S_{\mathrm{ISHE}}^{\mathrm{SSE}} T J_c - J_c^2 R_{\mathrm{ISHE}}^{\mathrm{SSE}}/2}.
\tag{88}
$$

The maximum efficiency of this power conversion scheme $\eta_{max}^{\mathrm{SSE}}$, (85) is reached at the optimal $J_c$:

$$
J_c^{opt} = \frac{S_{\mathrm{ISHE}}^{\mathrm{SSE}} \Delta T}{R_{\mathrm{load}}^{opt} + R_{\mathrm{ISHE}}}; \quad R_{\mathrm{load}}^{opt} = \frac{R_{\mathrm{ISHE}}^{\mathrm{SSE}}}{\sqrt{1+(ZT)_{\mathrm{ISHE}}^{\mathrm{SSE}}}}.
\tag{89}
$$

The different form for the optimal $R_{\mathrm{load}}$ compared to the conventional thermoelectrics (23) is again caused by the different sign of spin Seebeck and Peltier coefficients. The spin Seebeck figure of merit for the ISHE scheme is

$$
(ZT)_{\mathrm{ISHE}}^{\mathrm{SSE}} = \frac{\theta_{\mathrm{SH}}^2 \frac{G_N S_S^2 T}{G_K} \tanh^2 \frac{d_N}{2\lambda}}{\left(\coth \frac{d_N}{\lambda} + \frac{G_N}{G_S} + \frac{G_N S_S^2 T}{G_K}\right) \left[\frac{d_N}{\lambda}\left(\coth \frac{d_N}{\lambda} + \frac{G_N}{G_S}\right) + \theta_{\mathrm{SH}}^2 \left(1 + \frac{2G_N}{G_S}\tanh \frac{d_N}{2\lambda}\right)\right]}.
\tag{90}
$$

Eq. (90) can be maximized by the $d_N/\lambda$ that solves the equation

$$
\frac{d_N^{opt}}{\lambda} + \theta_{\mathrm{SH}}^2 \approx \frac{1}{2} \sinh \frac{d_N^{opt}}{\lambda}.
\tag{91}
$$

Since $d_N > 0$ we should choose the positive solution of this equation (91). For larger $\theta_{\mathrm{SH}}$, optimal heat current conversion can be achieved by an increased $d_N/\lambda$. For the spin Hall angle of Pt $\theta_{\mathrm{SH}} = 0.1$ gives the optimal width $d_N^{opt} \approx 2.18\lambda$, leading to the figure of merit

$$
(ZT)_{\mathrm{ISHE}}^{\mathrm{SSE}} \approx \frac{4\theta_{\mathrm{SH}}^2 \frac{G_N S_S^2 T}{G_K} \exp\left(-\frac{d_N^{opt}}{\lambda}\right)}{\left(1 + \frac{G_N}{G_S}\right)\left(1 + \frac{G_N}{G_S} + \frac{G_N S_S^2 T}{G_K}\right)}.
\tag{92}
$$

$(ZT)_{\mathrm{ISHE}}^{\mathrm{SSE}}$ monotonically increases with $G_S/G_N$ and $G_N S_S^2 T/G_K$, see Fig. 13 and is strongly reduced by the exponential factor. The maximum value $\max[(ZT)_{\mathrm{ISHE}}^{\mathrm{SSE}}] = 4\theta_{\mathrm{SH}}^2 \exp(-d_N^{opt}/\lambda)$ is reached at $G_S \gg G_N$ and $T \to \infty$.

The spin conductance $G_S = L_S/S_S$ of the YIG|Pt interface can be estimated as $G_S/(wl) \sim 6 \times 10^{13} \ \Omega^{-1}\mathrm{m}^{-2}$, which is much smaller than the spin conductance of Pt $G_N/(wl) = 10^{15} \ \Omega^{-1}\mathrm{m}^{-2}$ [66]. $G_S/G_N \to 0$ leads to the simplification

$$
(ZT)_{\mathrm{ISHE}}^{\mathrm{SSE}} \to 4\theta_{\mathrm{SH}}^2 e^{-d_N^{opt}/\lambda} \frac{L_S^2 T}{G_N} \frac{1}{G_K + G_S S_S^2 T}.
\tag{93}
$$

$(ZT)_{\mathrm{ISHE}}^{\mathrm{SSE}}$ does not depend on sample area. Substituting $L_S = G_S S_S$ into (93) we find for low temperatures $(ZT)_{\mathrm{ISHE}}^{\mathrm{SSE}} \propto (G_S/G_N) G_S S_S^2 T/G_K$, and at high temperatures $(ZT)_{\mathrm{ISHE}}^{\mathrm{SSE}} \propto G_S/G_N \to 0$. The value of $(ZT)_{\mathrm{ISHE}}^{\mathrm{SSE}}$ is estimated to be about $10^{-4}$ for $\theta_{\mathrm{SH}} = 0.1$ [8]. In spite of the small figure of merit, the device in Fig. (9) is attractive for applications, because of its simplicity while output voltage and power are proportional to $l$ and area $wl$, respectively.

### B. Detection by spin valve

Here we derive the figure of merit for the spin valve device described in Section III-D. The effective heat conductance

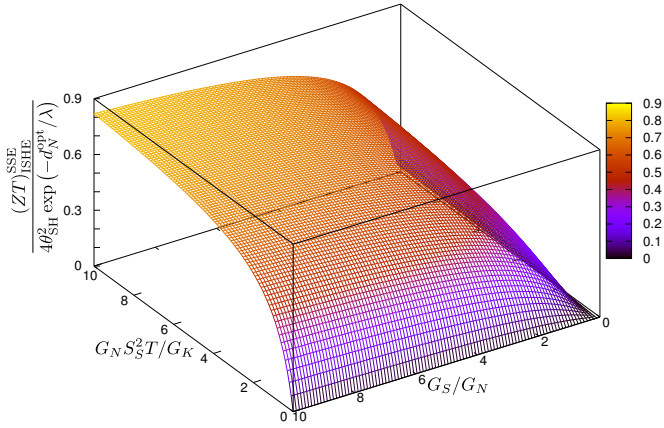

Fig. 13. Figure of merit $(ZT)^{\text{SSE}}_{\text{ISHE}}$ of a spin Seebeck power conversion device which utilizes the inverse spin Hall effect. $(ZT)^{\text{SSE}}_{\text{ISHE}}$ is a function of the ratio of the spin conductance and the electric conductance of N's magnetically active layer, $G_S/G_N$, and the dimensionless parameter $G_N S_S^2 T/K$. The magnitude of $(ZT)^{\text{SSE}}_{\text{ISHE}}$ is greatly enhanced when $G_S \gg G_N$ and $G_N S_S^2 T/K \to \infty$, however it is limited by the factor $\exp\left(-d_N^{\text{opt}}/\lambda\right)$.

$K^{\text{SSE}}_{\text{SV}}$ is obtained by substituting (68) into (83) and (84):

$$\Pi^{\text{SSE}}_{\text{SV}} = S^{\text{SSE}}_{\text{SV}} T \tag{94}$$

$$K^{\text{SSE}}_{\text{SV}} = G_K + S_S^2 T \frac{2(1-P^2)G_I G_S}{G_S + 2(1-P^2)G_I}. \tag{95}$$

The power conversion efficiency as a function of $J_c$ reads

$$\eta(J_c) = \frac{J_c S^{\text{SSE}}_{\text{SV}} \Delta T - J_c^2 R^{\text{SSE}}_{\text{SV}}}{K \Delta T + S^{\text{SSE}}_{\text{SV}} T J_c - J_c^2 R^{\text{SSE}}_{\text{SV}}/2}. \tag{96}$$

By setting the derivative with respect to $J_c$ to zero, we obtain the maximum efficiency $\eta^{\text{SSE}}_{lmax}$, Eq. (85), which is reached for

$$J_c^{opt} = \frac{S^{\text{SSE}}_{\text{SV}} \Delta T}{R^{opt}_{\text{load}} + R^{\text{SSE}}_{\text{SV}}}, \tag{97}$$

$$R^{opt}_{\text{load}} = R^{\text{SSE}}_{\text{SV}} \sqrt{1 + (ZT)^{\text{SSE}}_{\text{SV}}}, \tag{98}$$

with figure of merit

$$(ZT)^{\text{SSE}}_{\text{SV}} = \frac{2P^2 G_I S_S^2 T/G_K}{\left(1 + \frac{2G_I}{G_S}\right)\left[1 + (1-P^2)\left(\frac{2G_I}{G_S} + \frac{2G_I S_S^2 T}{G_K}\right)\right]}. \tag{99}$$

For $P = 0.8$ and 1, $(ZT)^{\text{SSE}}_{\text{SV}}$ is plotted in Fig. 14 as a function of $G_I/G_S$ and $G_I S_S^2 T/G_K$. The magnitude of $(ZT)^{\text{SSE}}_{\text{SV}}$ is greatly reduced as $G_I/G_S$ increases. The maximum value of $P^2/(1-P^2)$ is reached at $G_I \ll G_S$ and $T \to \infty$.

For intermetallic interfaces, $G_I/G_S \gg 1$ [20] and

$$\lim_{G_S/G_I \to 0} ZT^{\text{SSE}}_{\text{SV}} = \frac{P^2}{1-P^2} \frac{G_S}{2G_I} \frac{G_S S_S^2 T}{G_K}. \tag{100}$$

In the limit of a half-metal, this expression seems to diverge. However, but first letting $P \to 1$ and then $G_S/G_I \to 0$

$$\lim_{G_S/G_I \to 0} \lim_{P \to 1} ZT^{\text{SSE}}_{\text{SV}} = \frac{G_S S_S^2 T}{G_K}. \tag{101}$$

Estimating this value with the parameters of the ISHE-based device, we arrive at a value of 0.5 at room temperature. The scale independence of the spin valve charge current generation scheme might offer advantages for cooling of nanostructures.

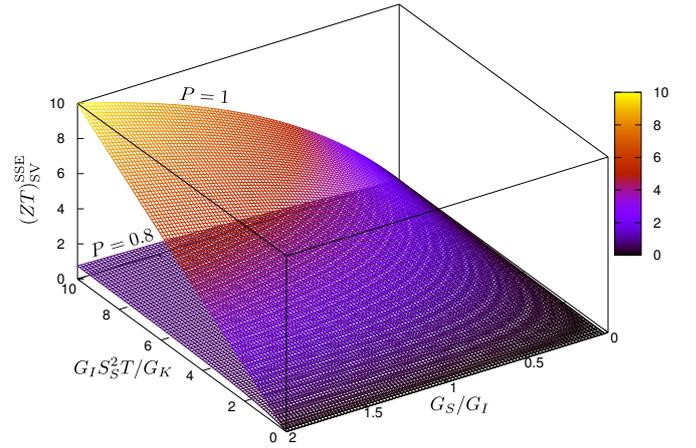

Fig. 14. The figure of merit of a spin Seebeck generator with spin valve conversion. $(ZT)^{\text{SSE}}_{\text{SV}}$ is a function of the ratio between the FI|N interface spin conductance $G_S$ and the N|FM electric conductance $G_I$ and the dimensionless parameter $G_I S_S^2 T/G_K$. $(ZT)^{\text{SSE}}_{\text{SV}}$ is enhanced when $G_S \gg G_I$ and $G_I S_S^2 T/G_K \to \infty$. The figure of merit $(ZT)^{\text{SSE}}_{\text{SV}}$ increases strongly with $P$ and plotted for $P = 1$ and $P = 0.8$.

## V. Conclusions

In this review we demonstrated that manipulating spin currents in thin-film multilayers and their nanostructures can improve the efficiency of heat to electric power conversion. This can be achieved by employing the spin dependence of the conventional thermoelectric coefficients in ferromagnets and their interfaces to normal metals. The discovery of the spin Seebeck and spin Peltier effects, generated by the collective spin wave excitations, opens entirely new strategies to enhance the conversion efficiency.

We considered two schemes for spin caloritronic power conversion devices based on spin Seebeck and spin Peltier effects. The SSE-ISHE device has a charge Peltier coefficient that has an opposite sign of its charge Seebeck coefficient, because of the anti-symmetric components in the spin Hall response matrix (65), while the SSE-spin valve obeys the conventional Onsager symmetry. By assuming that the heat conductance is limited by the Kapitza conductance of the FI|N interface, we are able to solve the power-conversion problem analytically.

The estimated figures of merit $(ZT)^{\text{SSE}}$ depend strongly on the spin-charge conversion scheme. The output voltage of the SSE-ISHE device is proportional to the sample length $l$ perpendicular to both the FI's magnetization and temperature gradient, Fig. 9, while the power scales with area, which is favorable for large area thermoelectric coatings, even though the figure of merit is greatly limited by the factor $\theta_{\text{SH}}^2 \exp\left(-d_N/\lambda\right)$ and spin Hall angles much smaller than unity.

Enhanced inverse spin Hall effect voltages were reported for magnetic multilayers subject to a temperature gradient as compared to the double layer. Lee et al. [36] reported that in CoFeB|Pt the anomalous Nernst and spin Seebeck effect enhance the voltage output. Ramos et al. [44] explained the observation that the output voltage increases with the number of periods in Fe$_3$O$_4$|Pt multilayers by an enhanced spin cur-

rents in superlattices. Both studies provide a new direction to improve the efficiency of large area spin caloritronic devices.

Employing spin valves to convert the spin accumulation into electric power promises much better figures of merit. The scale independence of the output voltage in the spin-valve setup can be useful for nanoscale applications, since the output voltage does not diminish when down-scaling the device. To increase the output voltage, a thermopile can be built based on the SSE-spin-valve setup. The utilization of spin valves to detect the SSE would further shed light on the role of interface proximity or spin-orbit interaction effects that might exist on YIG|Pt interfaces [30] but not on YIG|Cu [63] or YIG|Al [12]. The present modeling may also be applicable for other types of devices, such as spin Seebeck assisted magnetic random access memories [39].

### ACKNOWLEDGEMENTS

We thank E. Saitoh, K. Uchida, J. Flipse, and J. Xiao for insightful discussions. A.B.C. was supported by JSPS Fellowship for Young Scientists. We acknowledge support by the Grants-in-Aid for Scientific Research (Nos. 25800184, 25220910, 15H01009, and 25247056), the Dutch FOM/NWO, the German DFG via SPP 1538 "Spin Caloric Transport", EU-STREP InSpin, and DAAD SpinNet.